\documentclass[twocolumn,showpacs,prb]{revtex4}
\usepackage{graphicx}
\usepackage{dcolumn}
\usepackage{amsmath}
\usepackage{latexsym}
\begin{document}
%\wideabs{
\title{ A note on the noncommutative correction to gravity}
\author{ Pradip Mukherjee\footnote{Also Visiting Associate, S. N. Bose National Centre 
for Basic Sciences, JD Block, Sector III, Salt Lake City, Calcutta -700 098, India and\\ IUCAA, Post Bag 4, Pune University Campus, Ganeshkhind, Pune 411 007,India }$\,^{\rm a,b}$
Anirban Saha$\,^{\rm a,c}$\ }
\address{$^{\rm a}$Department of Physics, Presidency College\\
86/1 College Street, Kolkata - 700 073, India}
\address{$^{\rm b}$\tt pradip@bose.res.in}
\address{$^{\rm c}$\tt ani\_saha09@dataone.in}

%\address{Presidency College\\
%86/1 College Street, Calcutta -700 073, India.}

\begin{abstract}
{An apparent contradiction in the leading order correction to noncommutative (NC) gravity reported in the literature has been pointed out. We show by direct computation that actually there is no such controvarsy and all perturbative NC corrections start from the second order in the NC parameter. The role of symmetries in the vanishing of the first order correction is manifest in our calculation.}
\end{abstract}
%}
\maketitle
\section{Introduction}
The idea of fuzzy space time where the coordinates $x^{\mu}$ satisfy the
noncommutative (NC) algebra
\begin{equation}
\left[x^{\mu}, x^{\nu}\right] = i \theta^{\mu \nu}
\label{ncgeometry}
\end{equation}
with constant anti-symmetric $\theta^{\mu \nu}$, was mooted
long ago \cite{sny}. This idea has been revived in the recent past and
field theories defined over this NC space has been studied extensively \cite{szabo}. As always, a colossal challenge in this context comes from general theory of relativity (GTR). There are various attempts to fit GTR in the context of NC space time. Broadly there are two different approaches to the analysis of NC theories. One approach is to treat the fields as operators in some Hilbert space. The other approach is to define the fields over phase space with ordinary multiplication replaced by the Gronewald--Moyal product. In the later approach the original theory can be mapped to an equivalent commutative theory in the framework of perturbative expansion in the NC parameter, using the Seiberg--Witten-type maps \cite{sw, bichl} for the fields. This commutative equivalent approach has been used to analyse many gauge theories in the recent past \cite{ours}. Since gravity can be viewed effectively as a gauge theory the commutative equivalent approach seems to be a promising one
 . Indeed, a minimal theory of NC gravity \cite{cal} has been constructed recently based on this approach where the NC correction appears as a series expansion in the NC parameter. The leading order correction is reported to be linear in $\theta$ in this work.

  Construction of a theory of NC gravity remains a topic of considerable current interest in the literature and various authors have approached the problem from different angles. In \cite{Chamseddine:2000si} for example  a  deformation of Einstein's gravity was studied using a  construction  based on gauging the noncommutative SO(4,1) de Sitter group and the SW map \cite{sw} with subsequent contraction to ISO(3,1). Another construction of a noncommutative gravitational theory was proposed in \cite{Aschieri:2005yw}. Very recently noncommutative gravity has been connected with stringy perspective\cite{ag}. In all these works the leading order noncommutative effects appear in the second order in the NC parameter $\theta$.
  %Note that all these works are based on some deformation of the basic algebra.
  It seems thus that the result of \cite{cal} is in contradiction. In this brief report we extend their work to show that actually there is no such controversy.

 %A minimal formulation of Einstein's General Relativity on noncommutative spaces has been reported in \cite{cal} with undeformed algebra. This is based on the commutative equivalent approach where the original NC theory is mapped to a theory with commutative variables using SW-type map. the correction due to noncommutativity appears as a perturbative expansion in the NC parameter. Notably, here an order $\theta $ correction is shown which seems to contradict the other results \cite{Chamseddine:2000si, Aschieri:2005yw, ag}. The purpose of this comment is to show by direct calculation that there is no such controversy and the order $\theta $ minimal correction to gravity reported in \cite{cal} actually disappears. We first present a brief review of the results of \cite{cal}. This will be useful as the starting point of our calculation as well as to fix the notations.

 The organisation of this report is as follows. In the next section we present a brief review of the results of \cite{cal}. This will be useful as the starting point of our calculation as well as to fix the notations. In section 3 details of computation of the first order correction term has been given. Section 4 contains the concluding remarks.
%555555555555555555555555555555555555555555555555555555555555555555
 \section{Review of Minimal Formulation of NC Gravity}
%555555555555555555555555555555555555555555555555555555555555555555
 The main problem of implementing GTR on NC platform is that the algebra (\ref{ncgeometry}) is not invariant under general coordinate transformation.
However, we can identify a subclass of general coordinate transformations,
\begin{equation}
\hat x^{\mu \prime}=\hat x^{\mu}+\hat \xi^{\mu}(\hat x),
\label{c}
\end{equation}
which are compatible with the algebra given by (\ref{ncgeometry}). This imposes a restriction on $\xi^{\mu}$
\begin{eqnarray}
\theta^{\mu \alpha} \hat \partial_{\alpha} \hat  \xi^\nu(\hat x) = \theta^{\nu \beta} \hat \partial_{\beta} \hat \xi^\mu(\hat x).
\end{eqnarray}
and the theory corresponds to the version of General Relativity based on volume-preserving diffeomorphism known as the unimodular theory of gravitation \cite{UNI}. Thus the symmetries of canonical noncommutative space time naturally lead to the noncommutative version of unimodular gravity \cite{cal}. With the symmetries preserved in this manner the extension of GTR to noncommutative perspective is done using the tetrad formalism and invoking the enveloping algebra method \cite{Jurco:2000ja}. The theory is then cast in the commutative equivalent form by the use of appropriate Seiberg--Witten (SW) maps.
The final form of NC action is
\begin{eqnarray}
S = \int  d^4 x \frac{1}{2 \kappa^2} \hat R(\hat x)
\label{NCaction}
\end{eqnarray}
 $\hat R$ is the noncommutative version of the Ricci scalar
\begin{eqnarray}
\hat R = \hat R_{ab}{}^{ab} \label{R}
\end{eqnarray}
where $\hat {R}_{ab}{}^{cd}$ are the components of the NC Riemann
tensor appearing in
\begin{eqnarray}
\hat R_{ab}(\hat x) =\frac{1}{2}\hat R_{ab}^{\ \ cd}(\hat x)
\Sigma_{cd}, \label{Rab}
\end{eqnarray}
The latin indices refer to the vierbein and $\Sigma_{cd}$ are the
generators of the local Lorentz algebra $SO\left(3,1\right)$.

   $\hat R_{ab}(\hat x)$ can be expanded as \cite{cal}
\begin{eqnarray} \hat
R_{ab}=R_{ab} + R^{(1)}_{ab} + { \cal  O}(\theta^2)
\label{rab}
\end{eqnarray}
with
\begin{eqnarray}
R^{(1)}_{ab} =\frac{1}{2} \theta^{cd} \{R_{ac},R_{bd} \}
-\frac{1}{4} \theta^{cd} \{\omega_c, (\partial_d + D_d) R_{ab} \}.
\label{R1ab}
\end{eqnarray}
In the above expression $\omega_a^{\ bc}$ are the spin connection
fields
\begin{eqnarray}
\omega_a(x)=\frac{1}{2}\omega_a^{\ bc}\Sigma_{bc} \label{w}
\end{eqnarray}
which are antisymmetric under the exchange of $b$ and $c$. Again
$D_{a}$ is the covariant derivative
\begin{eqnarray}
D_a =  \partial_{a} + \frac{i}{2} \omega_a^{\ bc} \Sigma_{bc}
\label{covariant1}
\end{eqnarray}
 Note
that all quantities appearing on the rhs of (\ref{R1ab}) are
ordinary commutative functions.

  Using the above expansion we can write from (\ref{NCaction})
\begin{eqnarray}
S = \int  d^4 x \frac{1}{2 \kappa^2} \left (R(x)+
R^{(1)}(x)\right)+{\cal O}( \theta^2).
\end{eqnarray}
 $R(x)$ is the usual Ricci scalar and $R^{(1)}(x)$ is its
first order correction. In the following section we will
explicitly compute this correction term .
%________________________________________________________________________
\section{Explicit Computation of the First Order Correction Term}
%________________________________________________________________________
  For the computation of the first order term we need
an explicit form of $\Sigma_{cd}$, the generators of the local
Lorentz algebra $SO\left(3,1\right)$. This is given by
\cite{weinberg}
\begin{eqnarray}
\left[\Sigma_{cd}\right]^{a}{}_{b} = \delta^{a}{}_{c} \eta_{db} -
\delta^{a}{}_{d} \eta_{cb} 
\label{sigmat}
\end{eqnarray}
where $\eta_{ab} = \rm{diag}\left(-, +, +, +\right)$. As stated,
the latin indices refer to the vierbein. Our plan of calculation
is as follows. Using (\ref{sigmat}) we compute the first order
correction to the Ricci tensor (\ref{R1ab}). Since by finding the
$ab$-component of $\hat R_{ab}$ we get $\hat R$ on contraction,
the corresponding first order correction to the NC action can now
be calculated.

 We  now proceed to compute the correction term $R^{(1)}\left(x\right)$. First note that $R^{(1)}\left(x\right) = R^{\left(1\right)}_{ab}{}^{ab}$. From (\ref{Rab}) and (\ref{rab}) this is equal to  $\left[R^{\left(1\right)}_{ab}\right]^{ab}$. We thus have to calculate the corresponding matrix element of the rhs of (\ref{R1ab}) and contract. For convenience we write the result as
\begin{eqnarray}
\left[R^{\left(1\right)}_{ab}\right]^{ab}= T_{1} + T_{2}
\label{terms}
\end{eqnarray}
where  $T_{1}$ and $T_{2}$  denote respectively the contributions coming from the first and the second terms. Now after some computation we get
\begin{eqnarray}
T_{1} & = & 2 \theta^{cd} \left[R_{acg}{}^{a}R_{bd}{}^{bg} + R_{ac}{}^{b}{}_{g}R_{bd}{}^{ga}\right]\nonumber\\
%T_{2} & = & \frac{1}{2}4\theta^{cd} \left[\omega_{c}{}^{aj}\partial_{d}R_{abj}{}^{b} - \omega_{c}{}^{aj}\partial_{d}R_{ba}{}^{b}{}_{j}\right]
\label{T1}
\end{eqnarray}
The computation of the second term is somewhat involved. We first compute the part containing the covariant derivative as
\begin{eqnarray}
\left[\left(\partial_{d}+ D_{d}\right)R_{ab}\right]^{e}{}_{f} = 2\partial_{d}R_{ab}{}^{e}{}_{f} + i \omega_{d}{}^{eg}R_{abgf}
\label{covariant}
\end{eqnarray}
Using this expression for the derivative term we compute the second term to be
\begin{eqnarray}
T_{2} & = & -\theta^{cd} \left[\frac{1}{2}\left(\omega_{c}{}^{aj}\partial_{d}R_{abj}{}^{b} - \omega_{c}{}^{aj}\partial_{d}R_{ba}{}^{b}{}_{j}\right)\right.\nonumber\\ && \quad + \left. \frac{i}{4}\left(\omega_{c}{}^{aj}\omega_{dj}{}^{g}R_{abg}{}^{b} - \omega_{d}{}^{aj}\omega_{c}{}^{bg}R_{abjg}\right) \right]
\label{T2}
\end{eqnarray}
One can easily see that the first two terms cancel remembering the fact that the indices of the Riemann tensor refer to the tetrad and hence raised or lowered by $\eta_{ab}$. Collecting all the nonvanishing terms from (\ref{T1}, \ref{T2}) we get the correction term $R^{(1)}\left(x\right)$ as
\begin{eqnarray}
R^{(1)}\left(x\right)  &=&  \theta^{cd} \left[ 2\left(R_{acg}{}^{a}R_{bd}{}^{bg} + R_{ac}{}^{b}{}_{g}R_{bd}{}^{ga}\right)\right. \nonumber\\
&& \hspace{-2.50 mm} - \left. \frac{i}{4}\left(\omega_{c}{}^{aj}\omega_{dj}{}^{g}R_{abg}{}^{b} - \omega_{d}{}^{aj}\omega_{c}{}^{bg}R_{abjg} \right)\right]
\label{T}
\end{eqnarray}
This concludes our computation.
 Now one can show that all the terms of the above equation (\ref{T}) individually vanishes
 exploiting the antisymmetry of $\theta^{ab}$ and the various symmetry properties of the
 Riemann tensor and the spin connection fields.
%____________________________________________________
\section{Conclusion}
%____________________________________________________
Formulation of gravity in the perspective of noncommutative (NC) space time remains a topic of considerable current interest. There is an apparent contradiction in the results reported in the literature in the sense that where as in most of the works the noncommutative correction starts from the second order \cite{Chamseddine:2000si,Aschieri:2005yw,ag}, a minimal formulation of NC gravity \cite{cal} reports a first order correction. We have explicitly computed the first order correction term of the later work and demonstrated that it vanishes.
This has been shown to be due to the symmetries of the various
factors involved in the correction term. It appears  that in the perturbative framework the order $\theta$ correction must vanish because the zero order theory carries full local Lorentz symmetry.
\section{Acknowledgement}
 AS would like to thank the Council for Scientific and Industrial Research (CSIR),
 Govt. of India, for financial support.
  The authors also acknowledge the hospitality of IUCAA where part of the work
  has been done.\\

  {\bf{PS}} After the submission of our comment to the Physical
  review D there appeared a paper \cite{cal1} by the authors of \cite{cal}
  where they have given results of calculation to the second
  order. Naturally they have also found that the first order
  correction vanishes.

%%%%%%%%%%%%%%%%%%%%%%%%%%%%%%%%%%%%%%%%%%%%%%%%%%%%%%%%%%%%%%%%%
%%%
%%%                     BIBLIOGRAPHY
%%%%%%%%%%%%%%%%%%%%%%%%%%%%%%%%%%%%%%%%%%%%%%%%%%%%%%%%%%%%%%%%%%%%

\bigskip

%\newpage
%\vskip .75 in


\baselineskip=1.6pt
\begin{thebibliography}{99}

\bibitem {sny}Heisenberg first suggested this idea which was later developed by Snyder; H.~S.~Snyder, Phys. Rev. {\bf 71} (1947) 38; {\it{ibid}} 72 (1947) 874.

\bibitem{szabo} See R.~J.~Szabo, Phys. Rep. {\bf 378} (2003) 207 and the references therein.

\bibitem{sw}
N.~Seiberg, E.~Witten, JHEP {\bf 9909}, 032 (1999)
[hep-th/9908142].

\bibitem{bichl} A.~A.~Bichl, J.~M.~Grimstrup, L.~Popp, M.~Schweda, R.~Wulkenhaar, [hep-th/0102103]. 

\bibitem{ours}O.~F.~Dayi, B.~Yapiskann, JHEP {\bf{10}} (2002) 022, [hep-th/0208043]; 
S.~Ghosh, Nucl.Phys. {\bf B 670} (2003) 359, [hep-th/0306045]; B.~Chakraborty, S.~Gangopadhyay, A.~Saha, Phys. Rev. {\bf D 70} (2004) 107707, [hep-th/0312292]; S.~Ghosh, Phys.Rev.{\bf D70} (2004) 085007, [hep-th/0402029]; P.~Mukherjee, A.~Saha, Mod.Phys.Lett.{\bf{A21}} (2006) 821, [hep-th/0409248]; P.~Mukherjee, A.~Saha, [hep-th/0605123]; A.~Saha, A.~Rahaman, P.~Mukherjee, [hep-th/0603050].

\bibitem{cal} X.~Calmet, A.~Kobakhidze, Phys. Rev. {\bf D 72} 045010,2005, [hep-th/0506157]

\bibitem{Chamseddine:2000si}
  A.~H.~Chamseddine,  Phys.\ Lett.\ B {\bf 504}, 33 (2001)
 [hep-th/0009153].

\bibitem{Aschieri:2005yw} P.~Aschieri, C.~Blohmann, M.~Dimitrijevic, F.~Meyer, P.~Schupp, J.~Wess
 Class.Quant.Grav. {\bf 22} 3511 (2005), [hep-th/0504183]

\bibitem{ag} L.~Alvarez-Gaume, F.~Meyer, M.~A.~Vazquez-Mozo, [hep-th/0605113 ]

\bibitem{UNI}
J.~J.~van der Bij, H.~van Dam and Y.~J.~Ng, Physica {\bf 116A}, 307 (1982),
F.~Wilczek, Phys.\ Rept.\  {\bf 104}, 143 (1984);
W.~Buchmuller and N.~Dragon, Phys.\ Lett.\ B {\bf 207}, 292 (1988);
M.~Henneaux and C.~Teitelboim,Phys.\ Lett.\ B {\bf 222}, 195 (1989); W.~G.~Unruh,
 Phys.\ Rev.\ D {\bf 40}, 1048 (1989).

\bibitem{Jurco:2000ja}
B.~Jurco, S.~Schraml, P.~Schupp and J.~Wess, Eur.\ Phys.\ J.\ C {\bf 17}, 521 (2000),
[hep-th/0006246].

\bibitem{weinberg} S.~Weinberg, {\it {Gravitation and Cosmology: Principles and Applications opf the General Theory of Relativity}}, John Wiley \& Sons. , New York, 1972.
\bibitem{cal1} X.~Calmet, A.~Kobakhidze, [hep-th/0605275]

\end{thebibliography}
\end{document}